\documentclass[times,final]{elsarticle}
\journal{ Computational and Theoretical Chemistry}

\begin{document}

\begin{frontmatter}

\title{Weiner's theory for exactly solvable Schr\"odinger equation with symmetric double well potential}
\author{A.E. Sitnitsky}
\ead{sitnitsky@kibb.knc.ru}

\address{Kazan Institute of Biochemistry and Biophysics, FRC Kazan Scientific Center of RAS, P.O.B. 30,
420111, Russian Federation.}

\begin{abstract}
	The Weiner's theory (WT) is developed on the basis of the exactly solvable Schr\"odinger equation with trigonometric double-well potential (TDWP). The symmetric case of TDWP is considered. This modified version of WT (mWT) enables one to eliminate some severe approximations of the original Weiner's approach and to obtain more accurate results. An analytic formula is derived which provides the calculation of the proton transfer rate with the help of elements implemented in {\sl {Mathematica}}. We exemplify the application of mWT by calculating the proton transfer rate constant in the hydrogen bond of the  proton-bound ammonia dimer cation ${\rm{N_2H_7^{+}}}$ (${\rm{H_3N\cdot\cdot\cdot H^{+} \cdot\cdot\cdot NH_3}}$). The parameters of the model for this object are extracted from available literature data on IR spectroscopy and quantum chemical calculations. The approach yields the transition from the Arrhenius-like exponential temperature dependence characteristic of thermal activation to that of quantum tunneling. Besides it is well suited for describing the phenomenon of vibrationally enhanced tunnelling. 
\end{abstract}
\end{frontmatter}

\section{Introduction}
	Reaction rate theory is an old branch of chemistry and chemical physics which is still in active progress. There are many approaches to its constructing (transition state theory, Kramers' theory, Caldeira–Leggett model, etc). However many problems remain open especially in the field of enzymatic reactions. In particular the so-called phenomenon of vibrationally enhanced tunnelling (VET) is an interesting and tempting option for describing resonant processes in enzymatic hydrogen transfer (EHT) and in the polariton chemistry (see, e.g., a recent article of Hammes-Schiffer and coauthors \cite{Garn25} who pioneered this idea in enzymology and refs. therein). The implication of VET to EHT is a particular case of more general concept of dynamic mechanisms participating in reaction acceleration by enzymes (conventionally named "rate-promoting vibration"  or "gating") (see \cite{Wel86}, \cite{Bru92}, \cite{Ham95}, \cite{Sit95}, \cite{Bas99}, \cite{Koh99}, \cite{Ant01}, \cite{Al01}, \cite{Ag05}, \cite{Sit06}, \cite{Sit08}, \cite{Sit10}, \cite{Koh15}, \cite{Jev17} and refs. therein). It is based on the growing evidence of various dynamic modes in protein scaffold of an enzyme. Invoking the concept in the field of EHT is based on VET at proton transfer (PT) in hydrogen bonds (HB). It means the effect of resonant acceleration of the reaction by a coupled oscillation in some narrow frequency range \cite{Sok92}, \cite{Ham95}. For enzymes carrying out EHT there are few estimates of reaction acceleration (e.g., $10^{4}$ for alcohol dehydrogenase, $10^{12}$ for methylmalonylCoA mutase \cite{Sch09} and $10^{11}$ for triose-phosphate isomerase \cite{Silv21}, \cite{Alb76}, \cite{Alb76a}). The reason of such scarce experimental data is in considerable difficulties to model uncatalyzed reactions for such enzymes \cite{Sch09}. On the other hand theoretical modelling the phenomenon of VET poses challenges and  requires new tools even within the framework of highly elaborated approaches. The aim of the present article is to suggest the development of one of them specifically suited for the above mentioned purpose.

	The Weiner's theory (WT) \cite{Wei78}, \cite{Wei78a}, \cite{Wei81} is a well-established and fruitful model for the description of barrier crossing adiabatic chemical reactions. In contrast to the Caldeira–Leggett approach (which is based on a sound microscopic picture of the thermal bath) WT is from the outset built as a phenomenological generalization of the classical transition-state expression to the quantum regime. In it the notion of temperature is introduced at the thermal (to be specific Boltzmann) averaging of the flux and transmition coefficients associated with the right- and left-moving states. A similar approach is used in \cite{Gar14}. As mentioned above the main field of application for WT is the description of barrier crossing reactions. In particular PT in HB is a pattern process of barrier crossing (see, e.g.,  \cite{Gar14}, \cite{Jan73}, \cite{Sok92}, \cite{Ven99}, \cite{Ven01}, \cite{Shi11}, \cite{Yu16}, \cite{Nan25}, \cite{Gam22} and refs. therein).  Theretical modelling this process requires the solution of the Schr\"odinger equation (SE) with a double-well potential (DWP)  which is a one-dimensional cross-section of the potential energy surface \cite{Jan73}, \cite{Ven99}, \cite{Ven01}, \cite{Shi11}, \cite{Sit23}, \cite{Sit25}. DWP is one of the most important potentials in physics and chemistry  (see e.g., \cite{Jel12}, \cite{Tur16}, \cite{Ibr18}, \cite{Mar25} and refs. therein). There are many double-well potentials (DWP) for the Schr\"odinger equation (SE) but not all of them are convenient for applications in chemistry. For instance there is the famous double Morse potential belonging to a quasi-exactly solvable type for which exact expressions are derived via the functional Bethe ansatz for a limited number of energy levels. Unfortunately these expressions are very difficult for usage. Also there are DWPs (getting their names from the researchers as Rosen-Morse, Manning, Razavy, Konwent, etc) which do not allow obtaining the natural eigenfunctions for the Schrödinger’s operator. For this reason the authors expand the solution of SE over some suitable full sets of functions. The coefficients in the resulting series are determined from the appropriate recurrence relations. The latter makes the obtained solutions to be rather inconvenient for applications. For these reasons the corresponding SE are usually treated by quasi-classical (WKB) approximation or numerically with ensuing inconvenience in usage and scanning the parameter space of the model. Thus in the present article under exactly solvable DWPs we conceive those for which SE allows obtaining the natural eigenfunctions for the Schrödinger’s operator and as a result has a solution via an analytical special function with well-studied properties and preferably implemented in a generally accessible software.  Besides problems in chemistry usually require DWPs which take infinite values at the boundaries of the spatial variable interval. Finally it is desirable for DWP to be  smooth, i.e., not piecewise. Exact solutions for piecewise DWPs in terms of confluent hypergeometric functions (as, e.g., in the Quantum Mechanics textbook by Merzbacher) are of use for pedagogical purposes. However at practical applications the artificial construction of such DWPs casts some doubt upon the possible error introduced by sewing their parts. There are only two exactly solvable (in the above defined sense) DWPs (hyperbolic and trigonometric ones) satisfying the above mentioned requirements. In particular the necessities of the condensed matter physics stipulated the invention of exactly solvable hyperbolic DWPs (HDWP) which can be either finite at the boundaries of the spatial variable interval \cite{Xie12}, \cite{Dow13}, \cite{Che13}, \cite{Har14} or take infinite values there \cite{Don18}, \cite{Don181}, \cite{Don182}, \cite{Don183}, \cite{Don19}, \cite{Don191}, \cite{Don22}. For them such solutions of SE are feasible via the confluent Heun's function (CHF) implemented in {\sl {Maple}}. At approximately the same time a convenient trigonometric DWP (TDWP) was suggested for which SE has an exact solution via either CHF \cite{Sit17}, \cite{Sit171} or the spheroidal function (SF) \cite{Sch16}, \cite{Sit171}, \cite{Sit18} implemented in {\sl {Mathematica}}. SF is a well studied special function (see, e.g., the monograph \cite{Kom76}) and its spectrum of eigenvalues is also realized in {\sl {Mathematica}}. It extremely facilitates the calculation of the corresponding energy levels that makes TDWP to be highly suitable for chemical problems. Besides PT this TDWP was used in a number of other applications \cite{Sit19}, \cite{Sit20}, \cite{Cai20}, \cite{Sit21}, \cite{Cai22}. For HDWP obtaining energy levels is a challenging problem though it should be mentioned that a fine numerical algorithm was invented to overcome it (see \cite{Don18}, \cite{Don181}, \cite{Don182}, \cite{Don183}, \cite{Don19}, \cite{Don191}, \cite{Don22} and refs. therein). No applications of HDWP in chemistry have been published by now. In contrast for TDWP the solution is obtained via SF implemented in {\sl {Mathematica}} and energy leves are calculated easily with the help of the  spectrum of eigenvalues for this special function which is also realized in  {\sl {Mathematica}}. TDWP has various applications in chemistry by now. Its inherent limitation in describing PT in HB is in the fact that at present it can be conveniently used only for a symmetric potential (providing its parameters can be derived from experimental data on IR spectroscopy and/or quantum chemical calculations). The analytic solution for asymmetric TDWP is known via the generalized (Coulomb) SF \cite{Sit18}, \cite{Sit19} and the corresponding package for this function in {\sl {Mathematica}} was developed long ago by Falloon \cite{Fal01}. However it is unfortunately still not implemented in the standard versions of this software.  Anather limitation is in the requirement of only integer numbers for one of its two parameters $m$ (see below). They are related to two natural parameters (barrier height and barrier width) of a realistic DWP from data of quantum chemical calculations. However if such data provide more detailed information then there may appear difficulties in using only integer numbers for $m$.  This limitation can be eliminated if we make use of the solution of SE with TDWP via CHF which is also available. However in this case one encounters with above mentioned difficulties of obtaining energy levels in {\sl {Maple}}.

	In its original form WT is developed for DWP which has no exact analytic solution. As a result a number of severe approximations is introduced at deriving the probability flux and the quantum transmission coefficient. WT exists in two variants. One of them  \cite{Wei78} is based on the quantum scattering theory and treats PT in a rather stringent manner up to the moment when the details of DWP are needed. Then the author makes use of the piecewise quadratic DWP that leads to very cumbersome calculations due to the requirements imposed on wave functions and their derivatives by sewing different parts of the potential. Besides the artificial construction of this DWP casts some doubt upon the possible error introduced by such procedure. In the second variant \cite{Wei78a} Weiner makes use of the quasi-classical (WKB) method that enables one to circumvent the necessity of knowing the details of DWP but leads to its own difficulties. In particular the following approximations are invoked ($\psi_n \approx \psi_{n+1}$ to obtain eq. (3.3) and $ J_n \approx  J_{n+1}$;   $ P_n \approx  P_{n+1}$ to obtain the final result eq. (3.8) of \cite{Wei78a}). These approximations are very rough and can hardly be reliably justified. The aim of the present article is to show that TDWP enables one to eliminate these approximations of WT. We develop a modified WT (mWT) which is based on the scattering theory approach of WT. Making use of TDWP enables us to avoid the above mentioned troubles inherent for a piecewise DWP and to obtain more accurate results. Besides the above mentioned purely technical flaws original WT is known to have severe  conceptual problems \cite{Wei78}, \cite{Gar14}. For the piecewise quadratic DWP used in \cite{Wei78} the stationary states are expressed in terms of parabolic cylinder functions and the right- and left- moving states are individually nonnormalizable. As a result defining  the incoming and outgoing boundary conditions in the parabolic wells leads to asymptotically diverging solutions \cite{Gar14}. In our opinion using the uniform solution of SE with smooth (i.e., not piecewise) TDWP does not allow one to eliminate the problems of WT completely but at least alleviates the consequences of such drawbacks and difficulties of this theory. 

	To exemplify the general theory we consider PT rate for intermolecular HB in the proton-bound ammonia dimer cation
${\rm{N_2H_7^{+}}}$ (${\rm{H_3N\cdot\cdot\cdot H^{+} \cdot\cdot\cdot NH_3}}$). It is an important object in itself which is studied both by quantum chemistry \cite{Gar08} and IR spectroscopy \cite{Pri91}, \cite{Asm07}, \cite{Asa01}. Besides it is an interesting counterpart of the Zundel ion ${\rm{H_5O_2^{+}}}$ (oxonium hydrate ${\rm{H_2O\cdot\cdot\cdot H^{+} \cdot\cdot\cdot OH_2}}$) theoretically studied within the framework of WT in \cite{Sit23}, \cite{Sit25}. The Zundel ion at chosen there distances between oxygen atoms $R_{OO}=3.0\ \AA$ and $R_{OO}=2.8\ \AA$ has sufficiently high barriers to exclude the contribution of the over-barrier transition into PT rate constant even at high temperature. In contrast the ammonia dimer cation in our case of the distance between nitrogen atoms $R_{NN}=3.15\ \AA$ (literature data for the one-dimensional cross-section of the potential energy surface from quantum-chemical {\it ab initio} calculations are taken from \cite{Gar08}) is an example of a low-barrier HB. It provides vivid demonsration of the transition from the Arrhenius-like exponential temperature dependence characteristic of thermal activation to that of quantum tunneling.  Our results enable us to discuss the range of validity of the Goldanskii's criterion for the so-called "crossover" temperature $T_c$. Besides we briefly touch upon the phenomenon of VET.  Its consideration within the framework of mWT requires a 2D model but some important conclusions can be derived with the help of the present 1D model. The main aim of this article is to describe thoroughly mWT rather than its possible applications. For this purpose 1D model is quite sufficient and representable. We show that mWT provides a workable and self-consistent scheme for obtaining VET. The results obtained within the framework of mWT in our opinion are founded on firmer ground than those derived previously with the help of WT in \cite{Sit25} but in agreement with them and thus corroborate them.

	The paper is organized as follows. In Sec.2 the results for SE with TDWP are briefly summarized. In Sec.3 we derive a general expression for PT rate constant within the framework of mWT. In Sec.4 the results are discussed and the conclusions are summarized. In Appendix 1 the transformation of dimensional SE into its dimensionless form is discussed. In Appendix 2 some formulas of the original Weiner's theory are presented for the convenience of readers and the comparison of mWT and WT results is provided.  In Appendix 3 the list of abbreviations used is presented.\\

\section{Solution of Schr\"odinger equation}
We consider the one-dimensional SE with TDWP (\ref{eq2}) for the energy levels $\epsilon_q$ and the corresponding wave functions $\psi_q (x)$
\begin{equation}
\label{eq1}  \psi''_q(x)+\left[\epsilon_q-U(x)\right]\psi_q(x)=0
\end{equation}
Further we identify the normal mode under consideration with that of the proton in HB. Thus in (\ref{eq1}) $x$ is the proton coordinate. For the one-dimensional cross-section of the potential energy surface $U(x)$ we make use of TDWP \cite{Sit17}, \cite{Sit18}
\begin{equation}
\label{eq2} U(x)=\left(m^2-\frac{1}{4}\right)\ \tan^2 x-p^2\sin^2 x
\end{equation}
Here $-\pi/2 \leq x \leq \pi/2$, $m$ is an integer number and $p$ is a real number. The two parameters of TDWP $m$ and $p$ are related to two main characteristics of the potential energy surface, i.e., the barrier hight and the barrier width (see Appendix 1). The examples of TDWP for intermolecular HB in the proton-bound ammonia dimer cation is presented in Fig.1.
\begin{figure}
\begin{center}
\includegraphics* [width=\textwidth] {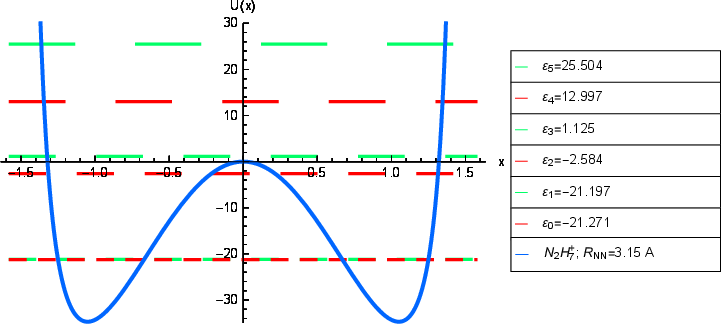}
\end{center}
\caption{The trigonometric double-well potential (\ref{eq2}) at the values of the parameters $m=2$; $p=7.82971$. The parameters are chosen to describe the hydrogen bond in the proton-bound ammonia dimer cation
${\rm{N_2H_7^{+}}}$  for the case $R_{NN}=3.15\ \AA$ (they are extracted from the data of quantum chemistry \cite{Gar08}). At the right of Fig.1 the energy levels (given by (\ref{eq4})) are presented.} \label{Fig.1}
\end{figure}
The exact analytic solution of (\ref{eq1}) is available \cite{Sit18}
\begin{equation}
\label{eq3} \psi_q (x)=\sqrt {\cos x}\ \bar S_{m(q+m)}\left(p;\sin x\right)
\end{equation}
where $q=0,1,2,...$ and $\bar S_{m(q+m)}\left(p;s\right)$ is the normalized angular prolate spheroidal function \cite{Kom76}. It is implemented in {\sl {Mathematica}}
as $\rm{SpheroidalPS}[(q+m),m,ip,s]$ where $i=\sqrt {-1}$ (note that the representation in this program software package is a non-normalized one). The energy levels are
\begin{equation}
\label{eq4} \epsilon_q=\lambda_{m(q+m)}\left(p\right)+\frac{1}{2}-m^2-p^2
\end{equation}
Here $\lambda_{m(q+m)}\left(p\right)$ is the spectrum of eigenvalues for $\bar S_{m(q+m)}\left(p;s\right)$. It is implemented in {\sl {Mathematica}} as $\lambda_{m(q+m)}\left(p\right)\equiv \rm{SpheroidalEigenvalue}[(q+m),m,ip]$. For TDWP the position of the minimum in the right well is defined by the requirement
\begin{equation}
\label{eq5} \cos\ x_{min}=\left[\frac{m^2-1/4}{p^2}\right]^{1/4}
\end{equation}

\section{Modified Weiner's theory}
	In WT \cite{Wei78}, \cite{Wei78a} (in what follows we use dimensionless values defined in Appendix 1) the particle position is described by the stationary one-dimensional SE (\ref{eq1}) with a symmetric DWP (it is TDWP $U(x)$ given by (\ref{eq2}) in our case) which has the solutions for the energy levels $ \epsilon_q$ and the corresponding wave functions $\psi_q (x)$ (in our case of TDWP they are known and given by (\ref{eq4}) and (\ref{eq3}) respectively). For any DWP its energy levels form doublets below the barrier top. The rate constant consists of the contribution from the tunneling process and that from the over-barrier transition. Concerning the former WT deals with two important values. The first one is the probability flux $ J_{q;r}$ to the right of particles in the left well when the particle is in the q-th state. The second one is the quantum transmission coefficient, i.e., the fraction of those right-moving particles which are transmitted to the right well $\mid T_q \mid^2$. These values are defined for the energy levels below the barrier top. We recall the derivation of the Boltzmann factor $\exp \left(E_b/k_BT\right)$ in the reaction rate constant due to over-barrier transition ($ \Theta \left(E-E_b\right) $ is the classical transition probability in this case)
\[
k=const\int_{0}^{\infty} dE\ \Theta \left(E-E_b\right) \exp \left(-E/k_BT\right) dE\propto \exp \left(-E_b/k_BT\right)
\]
where $\Theta \left(x\right)$ is the step function ($\Theta \left(x\right)=1$ for $x \geq 0$ and $\Theta \left(x\right)=0$ for $x < 0$).
By analogy for energy levels above the barrier top we set $ J_{q;r}\mid T_q \mid^2\equiv1$. Then the reaction rate constant is a result of Boltzmann averaging of the product $ J_{q;r}\mid T_q \mid^2$ 
\begin{equation}
\label{eq6}  k(\beta)=\left[\sum_{n=0}^{\infty}e^{-\beta  \epsilon_n}\right]^{-1} \left\{\sum_{q=0}^N e^{-\beta  \epsilon_q} J_{q;r} \mid T_q \mid^2+\sum_{m=N+1}^{\infty}e^{-\beta  \epsilon_m}\right\}
\end{equation}
where $q=0,1,2,..., N$ so that there are $N+1$ levels below the barrier top. The first sum in the curly brackets corresponds to the contribution due to the tunneling process in the reaction rate. It is over the energy levels below the barrier top for which the notions of $J_{q;r}$ and $ \mid T_q\mid^2$ have sense. The second sum in the curly brackets corresponds to the over-barrier transition and $\epsilon_{N+1}$ is the first energy level above the barrier top. 

	The original version of WT is developed for the case of a DWP for which the exact analytic solution of SE is unknown. In our case of TDWP the wave function $\psi_q (x)$ is known that leads to important consequences. Following \cite{Wei78} we make use of the scattering theory approach and write the decomposition of the wave function
\begin{equation}
\label{eq7} \psi_q (x)=F_q(x)\cos \chi_q (x)=\frac{1}{2}F_q(x)\exp\left(i \chi_q(x)\right)+\frac{1}{2}F_q(x)\exp\left(-i \chi_q(x)\right)\equiv \psi_{q;r} (x)+\psi_{q;l} (x)
\end{equation}
where we introduce the definition of the right-moving ($\psi_{q;r}$) and left-moving ($\psi_{q;l}$) states. At known $\psi_q (x)$ (given by  (\ref{eq3}) for TDWP) there is no need to represent this single function by two unknown functions $F_q(x)$ and $\chi_q (x)$  and to construct two separate equations for them as is done in WT. Moreover in the case of TDWP such approach inevitably leads to the appearance in the relationship for the flux of the integrals 
\begin{equation}
\label{eq8} \int_{0}^{x} \frac{dy}{\psi_q^2 (y)}
\end{equation}
which may become divergent due to zeros of $\psi_q (x)$. This difficulty plagues the following analysis within the framework of original WT. The approach developed below under the name mWT eliminates divergent integrals in the flux and replaces them by a $1/\psi_q (x)$ term in the expression for $\tan \chi_q(x)$ where it is harmless. It is achieved by a siutable choice of $\chi_q (x)$ by setting it from the relationship
\begin{equation}
\label{eq9}   \tan  \chi_q(x)=- \frac{\psi'_q (y)}{\psi_q (x)}\left[\frac{\alpha_q (x)}{\psi_q^2 (x)+\mu_q^2}\right]
\end{equation}
where $\mu_q^2$ is a constant and $\alpha_q (x)$ is an even function of $x$ to be defined below. It should be mentioned that at any $q$ the resulting $\chi_q(x)$ is an odd function of $x$ 
\begin{equation}
\label{eq10} \tan \chi_q(-x)=-\tan\chi_q(x) ;\ \ \ \ \ \ \ \ \ \ \ \ \ \ \ \ \ \ \chi_q(-x)=-\chi_q(x); 
\end{equation}
Indeed in our symmetric TDWP at even $q$ the function $\psi_q (x)$ is even while $\psi'_q (x)$ is odd and for odd $q$ it is vice-versa. We stress the oddness of $\chi_q(x)$ because it will be exploited below at calculating $T_q$. Actually the procedure of defining $\chi_q (x)$ from  (\ref{eq9}) means that (taking into account that $\psi_q (x)$ is the known function) we replace our unknown function $F_q(x)= \psi_q (x)/\cos \chi_q (x)$ by the new unknown function $\alpha_q (x)$.  Such procedure is valid if $\mu_q^2$ and $\alpha_q (x)$ are obtained below in a self-consistent manner.

	The probability flux  $J_{q;r}$ corresponding to the right-moving state $\psi_{q;r}$ is defined as usual in quantum mechanics
\begin{equation}
\label{eq11} J_{q;r}(x)=\frac{i}{2}\left[\psi_{q;r}\psi'^{*}_{q;r}-\psi^{*}_{q;r}\psi'_{q;r}\right]=\frac{1}{4}\chi'_q(x)F^2_q(x)=\frac{1}{4}\chi'_q(x)\frac{\psi^2_q (x)}{\cos^2 \chi_q(x)}=\frac{1}{4}\psi^2_q (x)\frac {d \tan \chi_q(x)}{dx}
\end{equation}
Thus the flux is not a constant as in WT but a function of $x$. In our opinion it is natural to define the flux in the right well corresponding to the reaction product (i.e., in the point $x_{min}$) $J_{q;r}\equiv  J_{q;r}\left(x_{min}\right)$ to be inserted in the expression for the reaction rate constant  (\ref{eq6}). From (\ref{eq9}) and (\ref{eq11}) with taking into account that from (\ref{eq1}) one has $\psi''_q(x)=-\left(\epsilon_q-U(x)\right)\psi_q(x)$ we obtain
\[
J_{q;r}\equiv  J_{q;r}\left(x_{min}\right)=\frac{1}{4\left(\psi_q^2 \left(x_{min}\right)+\mu_q^2\right)}\Biggl\{\alpha_q \left(x_{min}\right)\left(\psi'_q \left(x_{min}\right)\right)^2-
\]
\begin{equation}
\label{eq12}\psi_q \left(x_{min}\right)\psi'_q \left(x_{min}\right)\left[\alpha'_q \left(x_{min}\right)-\frac{2\psi_q \left(x_{min}\right)\psi'_q \left(x_{min}\right)\alpha_q\left(x_{min}\right)}{\psi_q^2 \left(x_{min}\right)+\mu_q^2}\right]+\alpha_q \left(x_{min}\right)\psi^2_q \left(x_{min}\right)\left(\epsilon_q-U\left(x_{min}\right)\right)\Biggr\}
\end{equation}
One can see that in the particular case $\psi_q \left(x_{min}\right) \rightarrow 0$ we obtain 
\begin{equation}
\label{eq13}  \lim_{\psi_q \left(x_{min}\right) \rightarrow 0} J_{q;r}\left(x_{min}\right)\approx\frac{\left(\psi'_q \left(x_{min}\right)\right)^2}{4}
\left[\frac{\alpha_q \left( x_{min}\right)}{\psi_q^2 \left(x_{min}\right)+\mu_q^2}\right]+O\left(\psi_q \left(x_{min}\right) \right)
\end{equation}
This form of the flux provides its sharp increase at $\psi_q^2 \left(x_{min}\right) \rightarrow 0$ if $\mu_q^2$ is sufficently small that is necessory for the phenomenon of VET.
The term $\mu^2_q$ (which may be small but not zero) eliminates possible divergencies of $ J_{q;r}$ due to zeros of $\psi_q\left(x_{min}\right)$ in the denominator.

	From  (\ref{eq7}) we have
\begin{equation}
\label{eq14}  \frac{F'_q(x)}{F_q(x)}=\frac{\psi'_q (x)}{\psi_q (x)}+\frac{ \tan  \chi_q(x)}{1+ \tan^2  \chi_q(x)}\frac {d \tan \chi_q(x)}{dx}
\end{equation}
For our symmetric TDWP we define the constants $\pm a_q$ (internal points on the barrier rather than external points on the outward wings of the potential) by the requirement
\begin{equation}
\label{eq15} U(\pm a_q)= \epsilon_q
\end{equation}
To find $\alpha_q (x)$ we need some additional assumption. We recall the WKB expression which is valid far from the turning point $a_q$, e.g., in the vicinity of $x_{min}$ (see, e.g., \cite{Lan74})
\begin{equation}
\label{eq16}
\psi^{WKB}_q (x)=\frac{const }{\left(\epsilon_q-U(x)\right)^{1/4}}\cos\left(\int_{a_q}^x dy\ \sqrt{\epsilon_q-U(y)}-\pi/4\right)
\end{equation}
Comparing it with (\ref{eq7}) we see that for WKB expressions 
\begin{equation}
\label{eq17} 
 F^{WKB}_q (x)=\frac{const}{\left(\epsilon_q-U(x)\right)^{1/4}}; \ \ \ \ \ \ \ \ \ \ \ \ \ \ \ \ \ \ \ \ \ \ \frac {d\chi^{WKB}_q(x)}{dx}=\sqrt{\epsilon_q-U(x)}
\end{equation}
there is a relationship 
\begin{equation}
\label{eq18} 
 \frac {dF^{WKB}_q(x)}{dx}\left[F^{WKB}_q (x)\right]^{-1}=-\ \frac {d^2 \chi^{WKB}_q(x)}{dx^2}\left[2 \frac {d\chi^{WKB}_q(x)}{dx}\right]^{-1}
\end{equation}
We assume that the same relationship takes place for our $F_q(x)$ and $\chi_q(x)$
\begin{equation}
\label{eq19} 
 \frac {dF_q(x)}{dx}\left[F_q (x)\right]^{-1}=-\ \frac {d^2 \chi_q(x)}{dx^2}\left[2 \frac {d\chi_q(x)}{dx}\right]^{-1}
\end{equation}
Substituting it into (\ref{eq14}) we obtain a cumbersome relationship in which (in the vicinity of $x_{min}$) the term grouped at $\psi'_q (x)$ provides us with an equation for $\alpha_q (x)$
\begin{equation}
\label{eq20} \frac {d^2\alpha_q (x)}{dx^2}-\frac{2\left(\psi'_q \left(x_{min}\right)\right)^2}{\psi_q^2\left(x_{min}\right)+\mu_q^2}\alpha_q (x)=0
\end{equation}
Its solution with an appropriately chosen boundary condition yields
\begin{equation}
\label{eq21} \alpha_q (x)=C\left[\exp \left(\frac{\sqrt{2}\mid \psi'_q \left(x_{min}\right)\mid x_{min}}{\sqrt{\psi_q^2\left(x_{min}\right)+\mu_q^2}}\right)\right]^{-1}\exp \left(\frac{\sqrt{2}\mid \psi'_q \left(x_{min}\right)\mid \mid x\mid}{\sqrt{\psi_q^2\left(x_{min}\right)+\mu_q^2}}\right) ;\ \ \alpha_q \left(x_{min}\right)=C;\ \ \alpha'_q \left(x_{min}\right)=\frac{C\sqrt{2}\mid \psi'_q \left(x_{min}\right)\mid }{\sqrt{\psi_q^2\left(x_{min}\right)+\mu_q^2}}
\end{equation}
Here $C$ is an arbitrary constant and in our theoretical consideration we further set it to be $C=1$.  However for comparison of the dimensionless theoretical PT rate constants with some experimental dimensional ones it may become necessary to choose an another value (see Appendix 1). The term grouped at $\psi_q (x)$ in the above mentioned relationship at $x_{min}$ yields a cubic equation for obtaining $\mu_q^2$
\[
\left[\epsilon_q-U(x_{min})\right]\left\{2\left(\psi_q^2\left(x_{min}\right)+\mu_q^2\right)^{3/2}-3\sqrt {2}\psi_q\left(x_{min}\right)\left(\psi_q^2\left(x_{min}\right)+\mu_q^2\right)\right\}+
\]
\begin{equation}
\label{eq22}  2\left(\psi'_q \left(x_{min}\right)\right)^2\left\{\left(\psi_q^2\left(x_{min}\right)+\mu_q^2\right)^{1/2}-2\sqrt {2}\psi_q\left(x_{min}\right)\right\}=0
\end{equation}
At known $\epsilon_q$ and $\psi_q(x)$ (as it is in our case) this equation is easily solved by {\sl {Mathematica}}. Thus the discussed above additional assumption obtained with the hint from WKB (\ref{eq18}) provides us with a closed and self-consistent procedure for calculating $ \alpha_q \left(x_{min}\right)=1$, $\alpha'_q \left(x_{min}\right)$ and $\mu_q^2$ to be inserted in  (\ref{eq12}).

	To find the value $T_q$ for the transmission coefficient $\mid T_q \mid^2$ (and hence that $R_q=1-T_q$ for the reflection coefficient) we following \cite{Wei78} introduce the scattaring states $\phi_r$ and $\phi_l$ by definitions
\begin{equation}
\label{eq23} \phi_{q;r}^{(1)}(x)=\psi_{q;r}(x)+R_q\psi_{q;l}(x)\ \ \ \ \ \ \ \ \ \ \ \ \ \ \ \ \ \ x \leq -a_q
\end{equation}
\begin{equation}
\label{eq24} \phi_{q;l}^{(1)}(x)=T_q\psi_{q;l}(x)\ \ \ \ \ \ \ \ \ \ \ \ \ \ \ \ \ \ \ \ \ \ \ \ \ x \leq -a_q
\end{equation}
\begin{equation}
\label{eq25} \phi_{q;r}^{(3)}(x)=T_q\psi_{q;r}(x)\ \ \ \ \ \ \ \ \ \ \ \ \ \ \ \ \ \ \ \ \ \ \ \ \ x \geq a_q
\end{equation}
\begin{equation}
\label{eq26} \phi_{q;l}^{(3)}(x)=\psi_{q;l}(x)+R_q\psi_{q;r}(x) \ \ \ \ \ \ \ \ \ \ \ \ \ \ \ \ \ \ x \geq a_q
\end{equation}
\begin{equation}
\label{eq27} \phi_{q;r}^{(2)}(x)=A_1\psi_q (x)+A_2\psi_q (-x) \ \ \ \ \ \ \ \ \ \  \ \ \ \ -a_q\leq x \leq a_q
\end{equation}
\begin{equation}
\label{eq28} \phi_{q;l}^{(2)}(x)=B_1\psi_q (x)+B_2\psi_q (-x)\ \ \ \ \ \ \ \ \ \ \ \ \ \ \ -a_q\leq x \leq a_q
\end{equation}
The constants $T_q$ ($R_q=1-T_q$), $A_1$, $A_2$, $B_1$ and $B_2$ are determined by the requirements that the scattaring states and their derivatives are continuous at $x=\pm a_q$, i.e., $\phi_{q;r}^{(1)}(-a_q)=\phi_{q;r}^{(2)}(-a_q)$, $\left[\phi_{q;r}^{(1)}\right]'(-a_q)=\left[\phi_{q;r}^{(2)}\right]'(-a_q)$ etc. As a result we have a system of algebraic equations and after cumbersome but straightforward calculations we obtain  
\begin{equation}
\label{eq29} T_q=\frac{\left[\psi_q (a_q)+\psi_q (-a_q)\right]\psi'_q (-a_q)-\left[\psi'_q (a_q)+\psi'_q (-a_q)\right]\psi_q (-a_q)}{\left[\psi_{q;r}(a_q)-\psi_{q;l}(-a_q)\right]\left[\psi'_q (a_q)+\psi'_q (-a_q)\right]-\left[\psi'_{q;r}(a_q)-\psi'_{q;l}(-a_q)\right]\left[\psi_q (a_q)+\psi_q (-a_q)\right]}
\end{equation}
For odd $q=2n+1$ (n=0,1,2,...,N) we have $\psi_{2n+1} (-a_q)=-\psi_{2n+1} (a_q)$, $\psi'_{2n+1} (-a_q)=\psi'_{2n+1} (a_q)$ and 
\begin{equation}
\label{eq30} T_{2n+1} =\frac{\psi_{2n+1}(a_{2n+1})}{\psi_{2n+1;r}(a_{2n+1})-\psi_{2n+1;l}(-a_{2n+1})}
\end{equation}
For even $q=2n$ (n=0,1,2,...,N) we have $\psi_{2n} (-a_{2n})=\psi_{2n} (a_{2n})$, $\psi'_{2n} (-a_{2n})=-\psi'_{2n} (a_{2n})$ and 
\begin{equation}
\label{eq31} T_{2n} =\frac{\psi'_{2n}(a_{2n})}{\psi'_{2n;r}(a_{2n})-\psi'_{2n;l}(-a_{2n})}
\end{equation}

	Knowing $\chi_q(x)$ and taking into account  (\ref{eq7}) we obtain after cumbersome but straightforward calculations
\begin{equation}
\label{eq32}  \mid T_{2n+1}\mid^2 =T_{2n+1}T_{2n+1}^*=\frac{1}{1+\left( \tan  \chi_{2n+1}(a_{2n+1})\right)^2}
\end{equation}
Analogous consideration for $T_{2n}$ yields
\begin{equation}
\label{eq33} \mid T_{2n}\mid^2 =\frac{\left[\psi'_{2n}(a_{2n})\right]^2\left[1+\left( \tan  \chi_{2n}(a_{2n})\right)^2\right]}{\left\{\psi'_{2n}(a_{2n})\left[1+\left( \tan  \chi_{2n}(a_{2n})\right)^2\right]+\psi_{2n}(a_{2n})\tan  \chi_{2n}(a_{2n})\tan'  \chi_{2n}(a_{2n})\right\}^2+\psi^2_{2n}(a_{2n})\left( \tan'  \chi_{2n}(a_{2n})\right)^2}
\end{equation}
In (\ref{eq32}) and (\ref{eq33}) the value $ \tan  \chi_q(x)$ is given by (\ref{eq9}) in which $\mu_q^2$ and $\alpha_q (x)$ are obtained from (\ref{eq22}) and (\ref{eq21}) respectively. As a result we express the desired combination $J_{q;r}\mid T_q\mid^2$ via the calculable values of our known function $\psi_q (x)$ given by (\ref{eq3}) and the energy level $\epsilon_q$ given by (\ref{eq4}) implemented in {\sl {Mathematica}}. We introduce the dimensionless inverse temperature $\beta$ (see Appendix 1). Substitution of the above obtained expressions into (\ref{eq6}) yields a workable expression easily programmed in this software\\
\newpage
\[
k(\beta)\approx\Biggl\{\sum_{m=M}^{\infty}e^{-\beta  \epsilon_m}+\frac{1}{4}\sum_{n=0}^N\frac{e^{-\beta  \epsilon_{2n+1}}}{\left[1+ \left( \tan  \chi_{2n+1}(a_{2n+1})\right)^2\right]\left(\psi_{2n+1}^2 \left(x_{min}\right)+\mu_{2n+1}^2\right)}\Biggl\{\alpha_{2n+1} \left(x_{min}\right)\left(\psi'_{2n+1}\left(x_{min}\right)\right)^2-\psi_{2n+1} \left(x_{min}\right)\times
\]
\[
\psi'_{2n+1} \left(x_{min}\right)\left[\alpha'_{2n+1} \left(x_{min}\right)-\frac{2\psi_{2n+1} \left(x_{min}\right)\psi'_{2n+1} \left(x_{min}\right)\alpha_{2n+1}\left(x_{min}\right)}{\psi_{2n+1}^2 \left(x_{min}\right)+\mu_{2n+1}^2}\right]+\alpha_{2n+1} \left(x_{min}\right)\psi^2_{2n+1} \left(x_{min}\right)\left(\epsilon_{2n+1}-U\left(x_{min}\right)\right)\Biggr\}+
\]
\[
\frac{1}{4}\sum_{n=0}^N\Biggl\{\frac{e^{-\beta  \epsilon_{2n}}\left[\psi'_{2n}(a_{2n})\right]^2\left[1+\left( \tan  \chi_{2n}(a_{2n})\right)^2\right]}{\left\{\psi'_{2n}(a_{2n})\left[1+\left( \tan  \chi_{2n}(a_{2n})\right)^2\right]+\psi_{2n}(a_{2n})\tan  \chi_{2n}(a_{2n})\tan'  \chi_{2n}(a_{2n})\right\}^2+\psi^2_{2n}(a_{2n})\left( \tan'  \chi_{2n}(a_{2n})\right)^2}\Biggr\}\times
\]
\[
\Biggl\{\alpha_{2n} \left(x_{min}\right)\left(\psi'_{2n}\left(x_{min}\right)\right)^2-\psi_{2n} \left(x_{min}\right)\psi'_{2n} \left(x_{min}\right)\left[\alpha'_{2n} \left(x_{min}\right)-\frac{2\psi_{2n} \left(x_{min}\right)\psi'_{2n} \left(x_{min}\right)\alpha_{2n}\left(x_{min}\right)}{\psi_{2n}^2 \left(x_{min}\right)+\mu_{2n}^2}\right]+
\]
\[
\alpha_{2n} \left(x_{min}\right)\psi^2_{2n} \left(x_{min}\right)\left(\epsilon_{2n}-U\left(x_{min}\right)\right)\Biggr\}\frac{1}{\left(\psi_{2n}^2 \left(x_{min}\right)+\mu_{2n}^2\right)}\Biggl\{\left(\psi'_{2n}\left(x_{min}\right) \right)^2-
\psi_{2n}\left(x_{min}\right)\psi'_{2n}\left(x_{min}\right)\times
\]
\begin{equation}
\label{eq34} \left[\frac{\alpha'_{2n}\left(x_{min}\right)}{\psi_{2n}^2 \left(x_{min}\right)+\mu_{2n}^2}-\frac{2\psi_{2n} \left(x_{min}\right)\psi'_{2n}\left(x_{min}\right)\alpha_{2n}\left(x_{min}\right)}{\left(\psi_{2n}^2\left(x_{min}\right)+\mu_{2n}^2\right)^2}\right]+\psi^2_{2n}\left(x_{min}\right)\left(\epsilon_{2n}-U\left(x_{min}\right)\right)\Biggr\}\Biggr\}\left[\sum_{q=0}^{\infty}e^{-\beta  \epsilon_q}\right]^{-1} 
\end{equation}
Here $2N$-th level is the last even one below the barrier top and $M$-th level is the first one above it. In Fig.1 an example of the case is presented that the last odd level of the doublets is above the barrier top so that $M=2N+1$. 

\section{Results and discussion}
We exemplify our model by its application to the ammonia dimer cation ${\rm{N_2H_7^{+}}}$ with the distance between nitrogen atoms $R_{NN}=3.15\ \AA$. This distance is the dimensional one $2L$ in our terminology (see Appendix 1). The results of the quantum-chemical {\it ab initio} calculations for the one-dimensional cross-section of the potential energy surface for this case are presented in Fig.1 of \cite{Gar08}. The barrier hight is $\Delta E \approx 1.3 \ kkal/mol$ and the distance of the minimum of the potential is $X_{min}=1\ \AA$. These data uniquely determine the parameter fitting for $p$ and $m$ by formulas in Appendix 1 providing the requirement for $m$ to be an integer number is satisfied. From the above data we obtain those of TDWP as $m=2$; $p=7.82971$. Generally (because of the requirement for $m$ to be an integer number) errors in $B$ and $W$ (from literature data of quantum chemical calculations) can certainly propagate to $m$ and $p$ and ultimately to PT rate constants. In our case the data on  quantum-chemical calculations provide a precise value for the barrier width but the boundary limits (i.e., the distance between nitrogen atoms $R_{NN}$) remain undefined. This enables us to choose the adjustable parameter $L=R_{NN}/2$ so that the requirement for $m$ to be an integer number can be strictly satisfied. In our approach we have to fit the distance $R_{NN}=3.15\ \AA$ (i.e., to choose $L=R_{NN}/2=1.575$) to obtain the integer number $m=2$. In fact this value can not be precisely determined from Fig.1 of \cite{Gar08}. Only $X_{min}=1\ \AA$ is a reliable value for the spacial coordinate so that the barrier width  $W=2 x_{min}=2\pi X_{min}/(2L)=1.995$ and it yields $R_{NN}=2 \pi/W\ \AA=3.15\ \AA$ providing the integer number $m=2$. If both spatial characteristics are given by data on  quantum-chemical calculations then the case is more restrictive and a systematic error analysis for the parameter extraction process is necessary. For the meagre literature data on ${\rm{N_2H_7^{+}}}$ from \cite{Gar08} the precedure of extracting $m$ and $p$ is unambiguous. For $m=2$; $p=7.82971$ there is a complete doublet (the ground state one) and the even level of the second doublet below the barrier top (see Fig.1). It means that in (\ref{eq37}) we have  $N=1$ and $M=3$.  In the calculations of the rate constant we also take into account with a safety margin of many levels above the barrier top (i.e., replace $\infty$ in (\ref{eq37}) by $M_{max}=10$) to make sure that the contribution of the over-barrier transition is taken into account adequately.  As an example of convergence test for the sum we present the calculations for the dependence of the proton transfer rate constant on $M_{max}$ in the proton-bound ammonia dimer cation ${\rm{N_2H_7^{+}}}$ with $m=2$; $p=7.82971$  at $\beta=0.0345$ as a set of pairs $\left\{M_{max}; k(0.0345)\right\}$:  $\left\{3;  0.848841\right\}$, $\left\{4; 0.931103\right\}$, $\left\{5; 0.984536\right\}$, $\left\{6; 1.0161\right\}$, $\left\{7; 1.03339\right\}$, $\left\{8; 1.04219\right\}$, $\left\{9; 1.04635\right\}$, $\left\{10; 1.04818)\right\}$. The potential in Fig.1 is an example of the low-barrier HB. 

In Fig.2 the dependence of the reaction rate constant on the inverse temperature is presented for various cases of barrier shapes. In the discussed above intermediate case $m=2$; $p=7.82971$ the even level $q=2$ of the second doublet is below the barrier top while the odd one $q=3$ is above it (see Fig.1). The model case $m=2$, $p=7$ corresponds to very low-barrier HB for which there is only one ground state doublet below the top. The model case $m=2$; $p=12$ corresponds to relatively high-barrier HB for which there are three complete doublets below the top. The problem of the transition from the Arrhenius-like exponential temperature dependence characteristic of thermal activation to that of quantum tunneling (low-temperature plateau) is of interest for a long time \cite{Ben93}, \cite{Ben94}. The transition is believed to take place in a rather narrow temperature range. In 1959 Goldanskii suggested a criterion for this so-called "crossover" temperature $T_c$
\begin{equation}
\label{eq35} 
 T_c=\frac{const\ \hbar}{dk_B}\sqrt{\frac{E_b}{2M}}
\end{equation}
where $E_b$ is the barrier hight, $d$ is its width corresponding to the zero-point energy in the initial state and const is a constant of order unity depending on the barrier shape (for a parabolic barrier it is $2/\pi$, for rectangular $1/2$, and for triangular $3/4$). This rough criterion was verified in numerous experiments and shown to provide qualitatively correct estimate. The sufficiently sharp transition from the exponential temperature dependence takes place via much weaker power-like one down to the low-temperature limit of the reaction rate constant. In this limit the rate constant becomes temperature-independent. The Goldanskii's criterion indicates the characteristic temperature of tunneling as the upper limit of exponential domination of tunnel transitions over Arrhenius ones. It states that at sufficiently low temperatures (when the population of thermally activated energy levels vanishes) the sole contribution to the reactive flux consists of tunneling from the ground state. Thus the approach takes into account the decisive role of zero-point vibrations in the appearance of the low-frequency limit in real conditions.  Literature experimental data (see \cite{Ben93}, \cite{Ben94} and refs. therein) testify that the transition takes place for many systems regardless of their barrier hight and the criterion roughly holds its validity. In our context of PT in HB it means that the transition occurs both for normal and for low-barrier HB.  and Fig.2 testify that mWT is indeed able to exhibit the phenomenon for various model cases from the extremely low-barrier one $m=2$, $p=7$ up to the high-barrier one $m=2$, $p=12$. Unfortunately we are not aware of experimental data for PT rate constant in ${\rm{N_2H_7^{+}}}$. In fact our calculations (presented in Fig.2 and Fig.3) testify that the notion of $T_c$ is a rather conventional one because the transition region highly depends on the barrier shape. It is more or less well defined for the sufficiently high barrier with $p=12$ but becomes a vague value for lower ones. In the latter case the transition takes place in a rather wide temperature range and $T_c$ at a large scale is ill-defined. However Fig.3 shows that even in this case at a smaller scale the transition region can be destinguished more or less definitely and provides a vivid picture of the transition from the Arrhenius temperature behavior to that of quantum tunneling. For $R_{NN}=3.15\ \AA$  ($p=7.82971$) our calculations yield $T_c\approx 60\ K$ (see Appendix 1) and from (\ref{eq35}) one can obtain $T_c\approx 50\ K$. Taking into account the considerable uncertainty of the coefficient const in (\ref{eq35}) we conclude that our result is in good agreement with the criterion in the range of the applicability of the notion of crossover temperature.
\begin{figure}
\begin{center}
\includegraphics* [width=\textwidth] {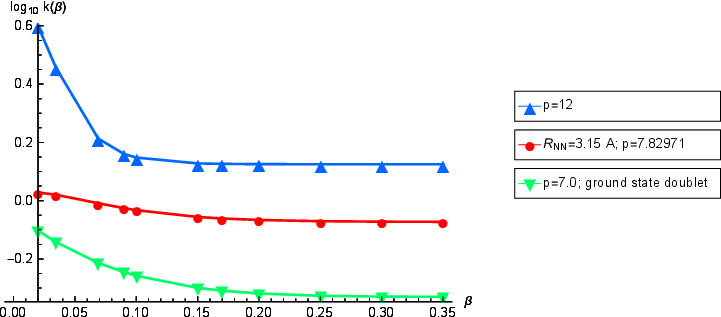}
\end{center}
\caption{The dependence of the proton transfer rate constant on the inverse temperature in the proton-bound ammonia dimer cation ${\rm{N_2H_7^{+}}}$  for the various cases of the parameter $p$ at $m=2$.} \label{Fig.2}
\end{figure}
\begin{figure}
\begin{center}
\includegraphics* [width=\textwidth] {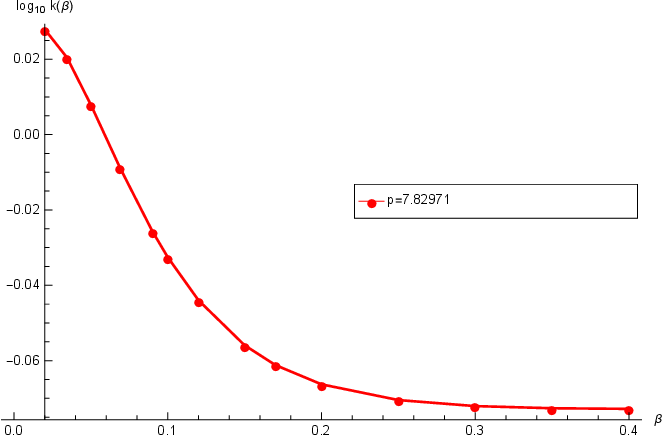}
\end{center}
\caption{The dependence of the proton transfer rate constant on the inverse temperature in the proton-bound ammonia dimer cation ${\rm{N_2H_7^{+}}}$  for the case $p=7.82971$ at $m=2$ corresponding to $R_{NN}=3.15\ \AA$.} \label{Fig.3}
\end{figure}
It is argued that $T_c$ derived from experimental curves are not in agreement with quantum tunneling calculations from one-dimensional models \cite{Ben93}, \cite{Ben94}.  These shortcomings of 1D models were attributed namely to their dimensionality and believed to be eliminated in multidimensional ones. In the latter case dynamic effects (named as VET or else vibration-assisted tunneling, vibrationally promoted one, etc) create qualitatively new features compared to the static barrier. They are actively studied since early 1980-s (see, e.g., \cite{Ben93}, \cite{Ben94}, \cite{Gri98}, \cite{Ahm25}, \cite{Garn25} and refs. therein). In our opinion the root of above mentioned troubles with 1D models is in the reaction rate theories invoked rather than in their dimensionality. mWT provides a rather stringent approach to the problem and our 1D model yields a broad range of possibilities to describe adequately experimental curves encountered in practice. It should be noted that WT for a 2D model with TDWP is available \cite{Sit23}, \cite{Sit25} and the phenomenon of VET was considered there. However in the present article we restrict ourselves by the 1D case and carry out only a brief survey of the problem within the framework of mWT. In fact the 2D model can be reduced to an effective 1D averaged SE \cite{Sit25}. In the case of coupling the reaction coordinate with an external oscillator the wave functions and the energy levels depend on its frequency $\omega$.  In the resulting approximate 1D SE they are replaced by their effective counterparts $ \psi_q (x) \longrightarrow \psi_q (x, \omega)$ and $\epsilon_q \longrightarrow \epsilon_q (\omega)$
\begin{equation}
\label{eq36}   \psi''_q (x, \omega)+\left[\epsilon_q(\omega)-U (x)\right]\psi_q (x, \omega)=0
\end{equation}
The form of the flux (\ref{eq12}) within the framework of our mWT is in a natural way compatible with the possibility of VET (in fact as was emhasized above it is specifically constructed for this purpose). The requirement for this phenomenon is a sharp increase of the flux.  At a peculiar resonant frequrency $\omega_{res}$ for some level (to be specific we denote it to be $q=2\bar n$) the condition may be realised that $\psi_{2\bar n}\left(x_{min}, \omega_{res}\right) \rightarrow 0$ and the flux $J_{2\bar n;r}$ (taking the limit form (\ref{eq13})) yields considerable growth. In Fig.4 we present the results for $J_{2;r}(\omega)$ ($\bar n=1$) calculated from corresponding data for the Zundel ion ${\rm{H_5O_2^{+}}}$ studied in \cite{Sit23}, \cite{Sit25}. In this case we have $\omega_{res}=0.01148149240963$, $\mu_2^2 \left(\omega_{res}\right)=2.471\cdot 10^{-21}$ and $\mid T_2\mid^2(\omega)=1$ at any frequency. We conclude that the flux indeed exhibits enormous resonant increase which provides the corresponding growth of the PT rate constant at the frequency $\omega_{res}$ by $26$ orders of magnitude compared to nonresonant plateau. One can consider the high potential barrier of HB in the Zundel ion as a rough model for those in EHT. Then our results suggest that reaction rate acceleration due to VET effect with a safety margin of sufficiency provides the values required for interpreting available experimental data in the field of EHT (see Introduction). Morover such reaction acceleration is commensurable with the examples of highest reaction acceleration in enzyme catalysis (the most efficient species do it up to factors more than $10^{20}$ \cite{Sch09}, e.g., $10^{21}$ for phosphohydrolases or $10^{26}$ for sulfate monoesters). The comment may be relevant because "poton transfer is the most common enzyme-catalyzed reaction, appearing in well over half of catalytic mechanisms" \cite{Silv21}. The result obtained within the framework of mWT is in agreement with those derived with the help of original WT in \cite{Sit25} but is more justified and rests on a firmer foundation. Thus mWT in our opinion provides a workable and self-consistent scheme for obtaining VET. 
\begin{figure}
\begin{center}
\includegraphics* [width=\textwidth] {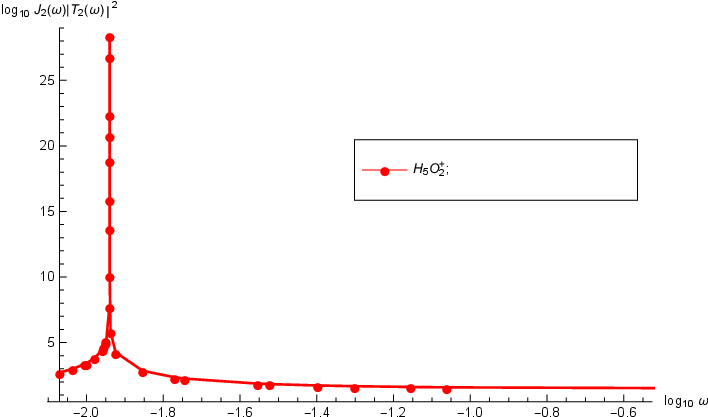}
\end{center}
\caption{The dependence of the probability flux and the quantum transmission coefficient product for the third level ($q=2$, i.e., $\bar n=1$) on the frequency of the external oscillator in the Zundel ion ${\rm{H_5O_2^{+}}}$ (oxonium hydrate) with $R_{OO}=3.0\ A$ (the values of the parameters for TDWP are $m=57$; $p=76$). The value of the coupling constant between the proton coordinate and that of the oscillator is $\alpha=5$ and the ratio of the mass for the proton to that of the oscillator is $\delta=1$  (see \cite{Sit25} for details). } \label{Fig.4}
\end{figure}

	 In the article we present the first application of mWT to HB in the proton-bound ammonia dimer cation and in the Zundel ion. However TDWP was applied earlier to HB in ${\rm{KHCO_3}}$ \cite{Sit17},  in chromous acid (CrOOH) and potassium dihydrogen phosphate (${\rm{KH_2PO_4}}$) \cite{Sit19} which are examples of intramolecular HB. This suggests that mWT can be applied to such systems as well. The main requirement for the applicability of mWT at present is the availability of reliable experimental data on IR spectroscopy and/or quantum chemical calculations for the corresponding DWP. Such data provide necessary adjustments to the model parameters $p$ and $m$ for different HB systems.

	 The suggested mWT enables one to eliminate the approximations of WT originated from the used in the original WT piecewise DWP. mWT deals with a uniform wavefunctions obtained for SE with the smooth TDWP. Besides TDWP is defined on a finite spacial interval and the exact analytic wave functions naturally decay at its boundaries. As a result one need not concern about asymptotic behavior of the wave functions at infinity as for DWP used in original WT with its parabolic wells. Finally WT treats complete doublets below the barrier top. The case of ${\rm{N_2H_7^{+}}}$ with $m=2$; $p=7.82971$ (for which the even level $q=2$ of the second doublet is below the barrier top while the odd one $q=3$ is above it) is problematic for WT. mWT treats separate levels and this case is of no difficulty for this theory. In our opinion mWT at least alleviates the consequences of the some drawbacks and difficulties of WT. However mWT does not allow one to eliminate the main conceptual problem of WT completely and unfortunately the crucial approximation remains to be unimproved. Namely the right-and left-moving states and the scattaring states in the decomposition of the wave function $ \psi_q (x)=\psi_{q;r} (x)+\psi_{q;l} (x)$ are individually nonnormalizable. Only $\psi_q (x)$ is a normalized function. As a direct consequence of it the requirement $R_q=1-T_q$ for the value $T_q$ in the transmission coefficient $\mid T_q \mid^2$ and that $R_q$ for the reflection coefficient $\mid R_q \mid^2$  is less physically justified than the standard $\mid R_q \mid^2=1-\mid T_q \mid^2$ condition.  As a matter of fact only the requirement $R_q=1-T_q$ is compatible with the main assumption of WT for the right-and left-moving states and the scattaring states $ \psi_q (x)=\psi_{q;r} (x)+\psi_{q;l} (x)=\phi_{q;r}(x)+\phi_{l;r}(x)$. The attempt to make use of $\mid R_q \mid^2=1-\mid T_q \mid^2$ requires a skewed decomposition of the wave function and the result will certainly be not the Weiner's theory which is in the title of the present article.

	The phenomenon of VET strongly depends on the shape of TDWP and on the type of the coupling between the proton coordinate and the external oscillator. In the present article we consider the case of the vibrational strong coupling motivated by the polariton chemistry. Its main feature  is that the strength of interaction is proportional to the square root of the frequency of the oscillator rather than to be a constant as for the case of symmetric coupling (see below). As a result this type takes into account a physical requirement that the strength of interaction should diminish with lowering the frequency. The interaction term in the hamiltonian is (see \cite{Sit25} for details)
\begin{equation}
\label{eq37}  
\alpha \sqrt \omega z \sin^2 x
\end{equation}
where $x$ is the proton coordinate, $z$ is that of the oscillator of frequency $\omega$ produced, e.g., by the cavity electromagnetic field and $\alpha$ is the coupling constant, i.e., the parameter characterizing the strength of interaction. At sufficiently high freqency of the oscillator (say $\omega > \omega_c$ where $\omega_c$ is some critical frequency) one can avarage the 2-d SE over the variable $z$. The effective averaged 1-d SE takes the form (see \cite{Sit25} for details)
\begin{equation}
\label{eq38}  
\Biggl\{\frac{\partial^2}{\partial x^2}+\epsilon_q(\omega)-\left(m^2-\frac{1}{4}\right)\ \tan^2 x+\left(p^2+\frac{\alpha c_q(\omega)}{\omega}\right)\sin^2 x\Biggr \}\psi_q (x, \omega)=0
\end{equation}
where (with the designation $\delta$ for the ratio of the reduced mass for the proton to that of the oscillator)
\[
 c_q(\omega)=\alpha\int_{-1}^{1}d\eta\ \eta^2\ \left[\bar S_{m(q+m)}\left(p;\eta\right)\right]^2+\sum_{l=0, l \neq q}^{\infty}\exp\Biggl\{-\frac{\alpha^2}{2\sqrt{2\delta}\omega^2}\Biggl(\int_{-1}^{1}d\eta\ \eta^2\ \left[\bar S_{m(l+m)}\left(p;\eta\right)\right]^2-
\]
\begin{equation}
\label{eq39}  \int_{-1}^{1}d\eta\ \eta^2\ \left[\bar S_{m(q+m)}\left(p;\eta\right)\right]^2\Biggr)^2\Biggr\}\int_{-1}^{1}d\eta\ \left[\gamma \eta+ \alpha\eta^2 \right]\bar S_{m(q+m)}\left(p;\eta\right)\bar S_{m(l+m)}\left(p;\eta\right)
\end{equation}
Its exact analytic solution for the wave function $\psi_q (x, \omega)$ is
\begin{equation}
\label{eq40} 
\psi_q (x, \omega)=\sqrt {\cos x}\ \bar S_{m(q+m)}\left(\sqrt {p^2+\frac{\alpha c_q(\omega)}{\omega}};\sin x\right)
\end{equation}
The energy levels are determined by the relationship
\begin{equation}
\label{eq41} 
\epsilon_q(\omega)=\lambda_{mq}\left(\sqrt {p^2+\frac{\alpha c_q(\omega)}{\omega}}\right)+\frac{1}{2}-m^2-p^2-\frac{\alpha c_q(\omega)}{\omega}
\end{equation}
For the second doublet of HB in the Zundel ion with $m=57$; $p=76$ there is a ratio $R=\alpha c_q(\omega_r)/\omega_r=422.4559341491942$ that provides the enormous increase of the flux $J_{2;r}$ at the resonant frequency $\omega_r$. As usually the dimensionless $\omega << 1$ the exponential term in the the expression for $ c_q(\omega)$ is small and one can approximately write $ c_q(\omega)\approx \alpha\ const$. From here we have a requierement for the coupling constant 
\begin{equation}
\label{eq42} 
 \alpha > \sqrt {\frac{R\omega_c}{const}}=\alpha_c
\end{equation}
Thus there is a limit value of $\alpha_c$ which is defined by the critical frequency $\omega_c$. Below this limit value it is unreasonable to consider the coupling constant because the whole theory based on the averaging of 2-d SE with the requirement $\omega > \omega_c$ is invalid there. But above the limit value there is no coupling strength threshold providing the requirement for $R$ is satisfied.  The dimensionless parameter of mass ratio $\delta$ and the dimensionless frequency $\omega$ are
\begin{equation}
\label{eq43} 
\delta=\frac{M}{\mu};\ \ \ \ \ \ \ \ \ \ \ \ \omega=\frac{4\sqrt {2} M L^2}{\sqrt {\delta}\hbar \pi^{3/2}}\Omega
\end{equation}
where $M$ is the mass the proton in HB and $\mu$ is the effective mass of the oscillator with  the frequency $\Omega$. In EHT the reaction coordinate of a corresponding enzyme is presumably coupled to a dynamical mode of its portein scaffold with $\Omega \geq 100\ cm^{-1}$. Thus with a safety margin of sufficiency one can assume $\Omega_c = 1\ cm^{-1}$. In this case we have $\alpha_c\approx4$ and $\omega_c\approx0.0075$. Our model values $\delta=1$, $\alpha=5$ yielding $\omega_r=0.0114815$ satisfy the above requirements. However constructing a physical model for $\alpha$ in EHT is an important but separate problem in itself which is beyond the scope of the present article.

	The coupling constant $\alpha$ in an actual HB system depends on the particular model of the external oscillator. For instance for HB in the Zundel ion  the proton is coupled with the internal stretching mode of the heavy atoms. For this object there are potential energy surfaces for several $R_{OO}$ as a result of quantum chemical {\it ab initio} calculations \cite{Yu16}, \cite{Xu18}. The Hamiltonian of the two-dimensional SE includes the spatial variable $Z$ (that for the reduced mass of the heavy atoms in HB which in the case of the Zundel ion is the O-O stretching mode) of the harmonic potential $\Omega Z^2/2$ with the frequency $\Omega$. In this case another type of coupling (named the symmetric one) is used $\lambda Z X^2$ (this case was proved in \cite{Jan73} to be pertinent for the Zundel ion). Here $\lambda$ is a dimensional coupling constant for the case of the symmetric mode coupling term and $\mu$ is the reduced mass of the heavy atoms in HB $A_1-H \cdot\cdot\cdot A_2$. In the case of the Zundel ion it is $\mu=M_O/2$. As a result $\delta=M/\mu=2M/M_O$. Taking the proton mass $M=1$ a.u. and that of the oxygen atom $M_O=16$ a.u. we have $\delta=1/8$. The dimensionless coupling constant $\alpha$ for the symmetric mode coupling and the frequency of the oscillator coupled to the proton coordinate (see \cite{Sit23} for details) are
\begin{equation}
\label{eq44} 
\alpha=\frac{2^6\lambda ML^5}{\hbar^2 \pi^5};\ \ \ \ \omega=\frac{4\sqrt{2M\mu}L^2\Omega}{\hbar \pi^2};\ \ \mu=\frac{M_1M_2}{M_1+M_2}
\end{equation}
For the cases of HB in the Zundel ion with $R_{OO}=2.5\ \AA$, $R_{OO}=2.6\ \AA$ and $R_{OO}=2.7\ \AA$ the authors of \cite{Jan73} provide the estimates of the dimensional coupling constant $\lambda$ ($a_{21}$ in their Table.1) as 0.1 a.u., 0.1 a.u. and 0.05 a.u. respectively. For the dimensional frequency $\Omega/2$ ($a_{02}$ in their Table.1) they present the value 0.039 a.u. for all three distances.  From here we obtain the value of $\alpha=0.6$ at $R_{OO}=2.5\ \AA$ and $\alpha=0.3$ at $R_{OO}=2.7\ \AA$. Unfortunately the frequency of the internal stretching mode of the heavy atoms in HB is not an experimentally adjustable parameter and corresponding data can not be used for a research of VET.

	We conclude that our approach (named here as modified Weiner's theory) enables one to obtain a reliable theoretical tool for describing the temperature dependence of proton transfer rate in hydrogen bonds. The approach is based on the exact analytic solution of the one-dimensional Schr\"odinger equation with the trigonometric double-well potential. We derive an analytically tractable expression for the proton transfer rate constant. As an example of applications it is used for the calculation of the proton transfer rate constant in the  intermolecular hydrogen bond of the proton-bound ammonia dimer cation ${\rm{N_2H_7^{+}}}$. The parameters of the model for ${\rm{N_2H_7^{+}}}$ are extracted from available literature data on IR spectroscopy and quantum chemical calculations. In particular the approach exhibits the transition from the Arrhenius-like exponential temperature dependence characteristic of thermal activation to that of quantum tunneling. Besides it is well suited for describing the phenomenon of vibrationally enhanced tunnelling. 

\section{Appendix 1}
We consider the dimensional form of stationary SE $\left(H-E\right)\psi(X)=0$ with the hamiltonian
\begin{equation}
\label{eq45}  H=-\frac{\hbar^2}{2M}\frac{\partial^2}{\partial X^2}+V(X)
\end{equation}
We choose the finite interval of the spatial variable $-L \leq X \leq L$ and assume $V(X)$ to be a DWP taking infinite values at its boundaries $X=\pm L$. We identify the normal mode $X$ with that of the proton in HB with the mass $M$. The dimensionless values for the coordinate $x$ and the potential $U(x)$ are introduced as follows
\begin{equation}
\label{eq46} x=\frac{\pi X}{2L};\ \ \ \ \ \ \ \ \  U(x)=\frac{8ML^2}{\hbar^2 \pi^2}V(X);\ \ \ \ \ \ \ \ \ \ \epsilon=\frac{8ML^2}{\hbar^2 \pi^2}E
\end{equation}
where $-\pi/2\leq x \leq \pi/2$.  As a result we obtain the dimensionless SE (\ref{eq1}). In the case of the trigonometric DWP (\ref{eq2})
the transformation formulas for the parameters $\{m,p\}$ into $\{B,W\}$ ($B$ is the barrier hight and $W$ is the barrier width) are \cite{Sit19}
\begin{equation}
\label{eq47}p=\frac{\sqrt {B}}{1-\left[\cos\left(W/2\right)\right]^2};\ \ \ \ \  \ \ \ \ \ \ \ \ \ \ \ \ \ \ \ \ \  m^2-\frac{1}{4}=\frac{B\left[\cos\left(W/2\right)\right]^4}{\left\{1-\left[\cos\left(W/2\right)\right]^2\right\}^2}
\end{equation}
The dimensionless inverse temperature $\beta$ in (\ref{eq6}) and (\ref{eq34}) is introduced as
\begin{equation}
\label{eq48}\beta=\frac{\hbar^2\pi^2}{8ML^2k_BT}
\end{equation}

	Unfortunately at present we are not aware of the experimetal data on PT for HB in ${\rm{N_2H_7^{+}}}$ and can not verify our theoretical assesments. Here we can suggest only an abstract scheme for obtaining dimensional values for PT rate constants.  The transform to the dimensional rate constant for a particular experiment in our opinion is carried out with the help of the following coefficient. The dimensionless rate constant is relatated to the corresponding dimensional one $\Gamma(T)$ by the empirical  coefficient $C$ (see  (\ref{eq21}) and text below). In our model calculations it was set $C=1$. Now we write $k\left(\beta, C \right)$ so that
\begin{equation}
\label{eq49}
 \frac{\Gamma(T)}{k\left(\beta, C \right)}=\frac{\hbar}{ML^2}\sim 10^{13}\ s^{-1}
\end{equation}
if we take the proton mass $M\sim 10^{-24}g$ and the distance $L\sim 1\ \AA$. One has to have at least one experimental value at reference temperature  $T_{ref}$ to determine $C$ from a fitting procedure
\begin{equation}
\label{eq50}
\Gamma \left(T_{ref}\right)=\frac{\hbar}{ML^2}k\left(\beta_{ref}, C \right)
\end{equation}
In the absence of such reference point we retain the value $C=1$. For ${\rm{N_2H_7^{+}}}$ we have $L\approx 1.57\ \AA$.  At room temperature $T\approx 300\ K$ we obtain $\beta=0.0724$. Our calculations yield $k(0.0724)\approx 0.973328$.  For the dimensional rate constant we obtain
\[
 \Gamma(300)=k(0.0724)\frac{\hbar}{ML^2}\approx 2.36\cdot10^{12}\ s^{-1}
\]
Analogously at $T=100\ K$ we have $\beta=0.217$ and $k(0.217)\approx 0.85471$ so that $ \Gamma(100)\approx 2.08\cdot10^{12}\ s^{-1}$. At $T=70\ K$ we have $\beta=0.31$ and $k(0.31)\approx 0.847$ so that $ \Gamma(70)\approx 2.057\cdot10^{12}\ s^{-1}$. 
 At $T=60\ K$ we have $\beta=0.362$ and $k(0.362)\approx 0.846$ so that $ \Gamma(60)\approx 2.054\cdot10^{12}\ s^{-1}$. 
 At $T=50\ K$ we have $\beta=0.434$ and $k(0.434)\approx 0.846$ so that $ \Gamma(50)\approx 2.054\cdot10^{12}\ s^{-1}$.  Thus the crossover temperature in our case is  $T_c=60\ K$. Small difference in the values of the rate constant at room temperature and at low ones is naturally explained by the fact that we consider the case of a low-barrier HB. Of course the real values of $\Gamma(T)$ should be corrected by the above mentioned procedure of choosing an actual $C$ from the experimetal data on PT for HB in ${\rm{N_2H_7^{+}}}$. However our conclusion on the value of $T_c=60\ K$ should hold notwithstanding such choice. The agreement of our estimate with that from the Goldanskii's criterion $T_c^G\approx50\ K$ suggests in favor of this conclusion.

In Fig.2 - Fig.5 we present not experimental poins for which error bars are natural and necessary but the results of computer calculations from an analytical expression. These calculations have hundred percent reproducibility and no error bars can be indicated in our case.

\section{Appendix 2}
In the Weiner's theory \cite{Wei78}, \cite{Wei78a} the particle position is described by the stationary one-dimensional SE with symmetric DWP $V(X)$ which has the solutions for the energy levels $E_q$ and the corresponding wave functions $\psi_q (X)$
\begin{equation}
\label{eq52}  \frac{\hbar^2}{2M}\psi''_q(X)+\left[E_q-V(X)\right]\psi_q(X)=0
\end{equation}
The rate constant consists of the contribution from the tunneling process. Concerning it the Weiner's theory deals with two important values. The first one is the probability flux to the right of particles in the left well when the particle is in the q-th state $J_q$. The second one is the quantum transmission coefficient, i.e., the fraction of those right-moving particles which are transmitted to the right well $\mid T_q \mid^2$. According to \cite{Wei78}, \cite{Wei78a} the reaction rate constant is a result of Boltzmann averaging of the product $J_q\mid T_q \mid^2$ calculated over the complete doublets below the barrier top
\begin{equation}
\label{eq53}  k=\left[\sum_{q=0}^{\infty}e^{-E_q/\left(k_BT\right)}\right]^{-1} \left\{\sum_{n=0}^N e^{-E_{2n}/\left(k_BT\right)} J_{2n} \mid T_{2n} \mid^2\right\}
\end{equation}
where $n=0,1,2,..., N\ $, $E_{2n}$ is the energy for the level $2n$ described by the wave function $\psi_{2n} (X)$. In the Weiner's theory the quantum transmission coefficient is calculated for the doublets which are counted by the even energy levels. For this reason $n$ is fixed to be even (see the text below the formula (3.1) in Sec.III of \cite{Wei78a}). The sum in the curly brackets corresponds to the contribution due to the tunneling process in the reaction rate. It is over the energy levels below the barrier top for which the notions of $ J_{2n}$ and $ \mid T_{2n} \mid^2$ have sense. In (\ref{eq43}) it is suggested by Weiner that the quantum transmission coefficient of the lower level in the doublet is determined by the splitting of the energy levels in it. Thus $N+1$ is the number of doublets below the barrier top and $E_{2N}$ is the lower energy level in the last doublet in this region. As a result only the sum over doublets (i.e., even levels $q=2n$) is left. The Weiner's theory is based on the quasi-classical approximation of the solution of SE \cite{Wei78a}
\begin{equation}
\label{eq54}  \psi_{2n} (x)=\frac{B_{2n}}{\sqrt {P_{2n}(x)}}\cos\frac{1}{\hbar}\left(\int_0^x d\xi\ P_{2n}(\xi) + S_{2n}\right)
\end{equation}
for $x\geq 0$.  The expression for  $\mid T_{2n}\mid^2$ follows from (3.8) and (2.17) of \cite{Wei78a}
\begin{equation}
\label{eq55}  \mid T_{2n} \mid^2=\left[\frac{E_{2n+1}-E_{2n}}{4\hbar J_{2n}}\right]^2
\end{equation}
The expression for $J_{2n}$ is given by (2.14) of \cite{Wei78a}
\begin{equation}
\label{eq56}  J_{2n}=\frac{B_{2n}^2}{4M}
\end{equation}
Substitution of the results into (\ref{eq53}) yields
\begin{equation}
\label{eq57}  k=\left[\sum_{q=0}^{\infty}e^{-E_q/\left(k_BT\right)}\right]^{-1}\left\{\frac{M}{4\hbar^2}\sum_{n=0}^N e^{-E_{2n}/\left(k_BT\right)} \frac{\left[E_{2n+1}-E_{2n}\right]^2}{B_{2n}^2}\right\}
\end{equation}
It should be stressed that the textbook quasi-classical representation of the wave function in the region $E > V(X)$ ($X > X^{(0)}$ where $X^{(0)}$ is the turning point) has the form (see, e.g., \cite{Lan74})
\begin{equation}
\label{eq58} \psi (x)=\frac{B}{\sqrt {P(x)}}\cos\left(\frac{1}{\hbar}\int_{X^{(0)}}^X d\xi\ P (\xi) -\pi/4\right)
\end{equation}
where
\begin{equation}
\label{eq59} P(X)=\sqrt{2M\left[E-V(X)\right]}
\end{equation}
The dimension of $P(X)$ is that of $\hbar/L$ where $L$ is a spacial value.  From  (\ref{eq58}) it follows that the dimension of $B$ is that of $\sqrt {P(x)}$ because the wave function $\psi (x)$ at dimensionless $x$ given by  (\ref{eq46})  is a dimensionless entity. As a result for $ J_{2n}$ in (\ref{eq56}) to have the dimension $s^{-1}$ (and as a consequence for the rate constant $k$  to have the dimension $s^{-1}$) one should actually conceive $M$ in (\ref{eq56}) and (\ref{eq57}) as $ML$.

	In the right-hand side well of TDWP ($0 \leq x$) there are two turning points $x_{2n}^{(0)}$ and $x_{2n}^{(1)}$ ($0<x_{2n}^{(0)}<x_{min}<x_{2n}^{(1)}<\pi/2$).
They are determined as the solutions of the equation
\begin{equation}
\label{eq60} \epsilon_{2n}=U(x)
\end{equation}
To make use of the Weiner's approach we need to represent our known wave function $\psi_{2n} (x)$ (given by (\ref{eq3}))  in the quasi-classical form. We write it as a mix of the Weiner's formula (\ref{eq47}) and the textbook one (\ref{eq48})
\begin{equation}
\label{eq61} \psi_{2n} (x)=\frac{B_{2n} }{\sqrt {P_{2n} (x)}}\cos\left(\int_{x_{2n}^{(0)}}^x d\xi\ P_{2n} (\xi) -\pi/f_{2n}\right)
\end{equation}
for $ x_{2n}^{(0)}< x <  x_{2n}^{(1)}$. Here $f_{2n}$ is an adjustable constant in WT. The WKB expression for $P_{2n}(x)$  (\ref{eq49}) in our notation of the dimensionless form is
\begin{equation}
\label{eq62}  P_{2n}^{WKB}(x)=\sqrt{\epsilon_{2n}-U(x)}
\end{equation}
For the bottom of TDWP ($x=x_{min}$) we have from here
\begin{equation}
\label{eq63} \psi_{2n} \left(x_{min}\right)=\frac{B_{2n} }{\left[\epsilon_{2n}-U \left(x_{min}\right)\right]^{1/4}}\cos\left(\int_{x_{2n}^{(0)}}^{x_{min}}d\xi\  \left[\epsilon_{2n}-U(\xi)\right]^{1/2} -\pi/f_{2n}\right)
\end{equation}
Thus finally we obtain for even $q$ ($q=2n$ where $n=0,1,2,...$)
\begin{equation}
\label{eq64} B_{2n}^2=\psi_{2n}^2 \left(x_{min}\right)\left[\epsilon_{2n}-U \left(x_{min}\right)\right]^{1/2}\cos^{-2}\left(\int_{x_{2n}^{(0)}}^{x_{min}}d\xi\  \left[\epsilon_{2n}-U(\xi)\right]^{1/2} -\pi/f_{2n}\right)
\end{equation}
where $\psi_{2n} (x)$ and $\epsilon_{2n}$ are given by  (\ref{eq3}) and  (\ref{eq4}) respectively.
As a result the desired product is
\[
J_{2n} \mid T_{2n} \mid^2=\left[ \epsilon_{2n+1}- \epsilon_{2n}\right]^2/B_{2n}^2=
\]
\begin{equation}
\label{eq65} \left[\epsilon_{2n+1}- \epsilon_{2n}\right]^2\psi_{2n}^{-2} \left(x_{min}\right)\left[\epsilon_{2n}-U \left(x_{min}\right)\right]^{-1/2}\cos^{2}\left(\int_{x_{2n}^{(0)}}^{x_{min}}d\xi\  \left[\epsilon_{2n}-U(\xi)\right]^{1/2} -\pi/f_{2n}\right)
\end{equation}
Thus we express the value $J_{2n} \mid T_{2n} \mid^2$ via the calculable values implemented in {\sl {Mathematica}}.\\

We introduce the dimensionless inverse temperature $\beta$ (see Appendix 1). Further we restrict ourselves to the case of the Boltzmann statistics. The generalization of (\ref{eq57}) for the rate constant with taking into account the possibility of the over-berrier transitions (see  (\ref{eq6}) and text above) is
\[
k(\beta)\approx\Biggl \{\sum_{n=0}^N\ e^{-\beta \left(\lambda_{m(2n+m)}\left(p\right)\right)}\Biggl[\left(\lambda_{m(2n+1+m)}\left(p\right) \right)-\left(\lambda_{m(2n+m)}\left(p\right)\right)\Biggr]^2\cos^{-1} x_{min}\ \bar S^{-2}_{m(q+m)}\left(p;\sin x_{min}\right)\times
\]
\[
\left[\left(\lambda_{m(2n+m)}\left(p\right)+\frac{1}{2}-m^2-p^2\right)-U \left(x_{min}\right)\right]^{-1/2}\cos^{2}\left(\int_{x_{2n}^{(0)}}^{x_{min}}d\xi\  \left[\left(\lambda_{m(2n+m)}\left(p\right)+\frac{1}{2}-m^2-p^2\right)-U \left(\xi\right)\right]^{1/2} -\pi/f_{2n}\right)+
\]
\begin{equation}
\label{eq66}\sum_{l=2N+2}^{\infty}\ e^{-\beta \left(\lambda_{m(l+m)}\left(p\right)\right)}\Biggr\}\left[\sum_{q=0}^{\infty}\ e^{-\beta \left(\lambda_{m(q+m)}\left(p\right)\right)}\right]^{-1}
\end{equation}
The sum over $n=0,1,2,..., N\ $ is that over $N+1$ doublets below the barrier top. 

\begin{figure}
\begin{center}
\includegraphics* [width=\textwidth] {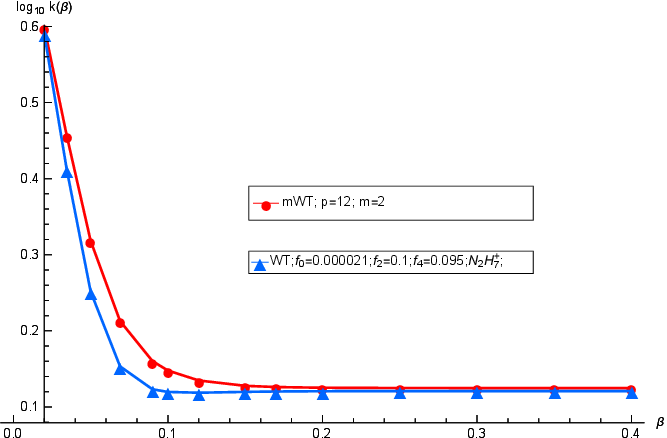}
\end{center}
\caption{The dependence of the proton transfer rate constant on the inverse temperature in the proton-bound ammonia dimer cation ${\rm{N_2H_7^{+}}}$  for the model case $p=12$ at $m=2$ calculated within the framework of the Weiners' theory (WT) and the  modified  Weiners' theory (mWT). The parameters of WT in (\ref{eq66}) are $f_0=0.000021$; $f_2=0.1$ and $f_4=0.095$.} \label{Fig.5}
\end{figure}

	In Fig.5 we present the comparison of mWT (\ref{eq34}) and WT (\ref{eq66}) results for the model case of the parameter $p=12$ at $m=2$. The realistic case of the proton-bound ammonia dimer cation ${\rm{N_2H_7^{+}}}$  for the case $p=7.82971$ at $m=2$ is problematic for WT because  (\ref{eq66}) is suited only for complete doublets below the barrier top and its unclear how to treat the splitted doublet in which the even $q=2$ is below it while the odd one $q=3$ is above it. One can see that there is good agreement between two theories in cases where WT can be applied. However conceptually mWT is more justified and contains fewer adjustable parameters than WT  (in our example $C=1$ in mWT versus  $f_0=0.000021$; $f_2=0.1$ and $f_4=0.095$ in WT). In our opinion the improvement in accuracy of mWT over WT can be established only by treating experimental data for some object. Unfortunately for ${\rm{N_2H_7^{+}}}$ such data are not available at present.

\section{Appendix 3}
Here we present the list of abbreviations used in the article.\\
VET- vibrationally enhanced tunneling\\
EHT - enzymatic hydrogen transfer\\
SE   - Schr\"odinger equation\\
WT - Weiners' theory\\
mWT - modified  Weiners' theory\\
HB   - hydrogen bond\\
PT   - proton transfer\\
DWP  - double-well potential\\
HDWP - hyperbolic double-well potential\\
TDWP - trigonometric double-well potential\\
WKB  - Wentzel-Kramers-Brillouin method or quasi-classical approximation\\
IR   - infra-red\\
CHF - confluent Heun's function\\
SF -   spheroidal function \\

\section*{Acknowledgments}
The author is grateful to Prof. Yu.F. Zuev for helpful discussions. The work was supported from the government assignment for FRC Kazan Scientific Center of RAS.

\end{document}